\documentclass[aps,amsmath,amssymb,amsfonts,twocolumn]{revtex4-1}
\usepackage[english]{babel}
\usepackage{mathtools}
\usepackage{physics}
\usepackage{xcolor}
\usepackage{graphicx}
\usepackage{lipsum}
\usepackage{comment}
\usepackage[normalem]{ulem}
\usepackage{hyperref}
\usepackage[utf8x]{inputenc}
\newcommand{\lx} {\left}
\newcommand{\rx} {\right}

\newcommand{\cA} {\mathcal{A}}

\newcommand{\cB} {\mathcal{B}}

\newcommand {\bfr} {\mathbf{r}}

\begin{document}

\title{Response of Worm-like Chains to Traveling-Wave Active Forces}

\author{Fabio Cecconi}
    \email[Correspondence email address:]{fabio.cecconi@cnr.it}
    \affiliation{Institute for Complex Systems - CNR, P.le Aldo Moro 2, 00185, Rome, Italy}
\author{Dario Lucente}
    \affiliation{Department of Mathematics~\& Physics - University of Campania “Luigi Vanvitelli”, Viale Lincoln 5, 81100 Caserta, Italy}
\author{Andrea Puglisi}
    \affiliation{Institute for Complex Systems - CNR, P.le Aldo Moro 2, 00185, Rome, Italy}
\author{Massimiliano Viale}
    \affiliation{Institute for Complex Systems - CNR, P.le Aldo Moro 2, 00185, Rome, Italy}

\date{\today} 

\begin{abstract}
We present a simplified theory for semiflexible flagella under the action of a traveling-wave perturbation that emulates the organized active forces generated by molecular motors, capable of inducing beating patterns to the filament.
By modeling the flagellum as a worm-like chain (WLC), we explore the interplay (competition) between the externally applied perturbation and the intrinsic bending rigidity of the filament. 
Our analysis aims to understand how this interplay can lead to a selection of conformations with the spatiotemporal behavior resembling the beating dynamics of axonemes such as those in sperm tails, in Chlamydomonas cilia, or eukaryotic flagella in general.
Through a systematic analysis of the WLC's response to traveling-wave perturbations, we try to identify the key parameters that mostly influence the flagellar motion and shape its beating-like profiles.
\end{abstract}


\maketitle

\section{Introduction}
Flagella are relatively slender structures found in various living organisms playing a crucial role in motility and locomotion. 
While bacterial flagella are usually passive filaments actuated by motors 
located at their base \cite{rev_BactMot}, eukaryotic flagella such as the cilia in organisms like Chlamydomonas or 
the tails of sperm cells are active filaments powered by numerous molecular motors distributed along their length \cite{gilpin2020multiscale}.

Flagella serve, in general, as the primary swimming apparatus that, by wiggling and rotating, enables microorganisms to move through fluid environments. 
Besides propulsion \cite{swimKeller}, they perform, along with cilia and bio-filaments, a variety of other biological processes \cite{khan2018assembly}, that include, for instance, movement of fluids and particles, pathogens removal, e.g. mucus clearance in lung tissue \cite{cleanAmJPhysiol,clearcilia}, embryogenesis (cell signaling, tissue development and patterning) \cite{arellano2012development,boehlke2010primary,wheway2018signaling}, mechanotransduction \cite{mechanotransduction} and sensory reactions to environmental stimuli \cite{lele2013dynamics}.

The complex dynamics of flagella have attracted researchers for decades, inspiring investigations into the fundamental mechanical principles that govern their biological function. 
In recent years, interest has also grown for the theoretical and experimental study of dynamical fluctuations in the flagellum beating pattern, which is periodic only on average \cite{ma2014active,maggi2023thermodynamic, sharma2024active}.
This research is driven by the natural need for knowledge but also by the goal of emulating their working efficiency in engineered micro-scale devices and macroscopic biomimetic systems \cite{mechanotransduction}.
Several theoretical approaches have been developed or applied to address the wide range of flagella shapes and their complex dynamics \cite{wang2017structural}, as well as to explain and interpret a plethora of experimental evidence and observations \cite{carroll2020flagellar,blair2003flagellar}.
Theory includes mechanical description based on the elasticity of flexible rod-like and slender bodies \cite{calldine1978change,hines1983three,moreau2018asymptotic, purohit2008mechanics,taloni2023general}, computational methods \cite{kitao2018molecular,schulten2006coarse,dillon2007fluid} and simple theoretical frameworks borrowed by active polymer physics \cite{zhu2024nonEq,ghosh_gov,actipoly,winkler2020physics}, each suited to account for the specific phenomenology of various types of flagella.
In particular, the bending rigidity of long linear biomolecules, flagella, and biofilaments in general, is a crucial characteristic that significantly assists their biological functions, and it is natural to assume that molecular stiffness is essential for the precise functional control of the beating dynamics.  
From this perspective, theoretical and computational models that accurately incorporate this bending rigidity are essential for capturing, mechanical, physical, and structural aspects of such functional dynamics.

Therefore, models of semiflexible polymers, which account for the stiffness of biopolymers like actin filaments, proteins, and DNA \cite{kas1996factin, ober2000shape}, are also useful for describing flagella. 
This stiffness arises from the valence angles in their backbone structure \cite{hofmann1997molecular}. 
However, these models are more difficult to handle theoretically, as they require additional constraints like preserving a fixed chain length.

A very popular approach resorts to employing the so-called  "worm-like chain" (WLC) models representing a minimal yet meaningful physical approach to implementing bending rigidity, as they describe flexible elongated structures characterized by a certain persistence length \cite{KPmodel,wlc_stat}. 
This property has strongly motivated the modeling of flagella as active persistent polymers \cite{actipoly,activesemiflex,ghosh_gov}, idealized 
as chains of units (molecular motors) that, through their organized or synchronized activity, determine the shape and the right collective dynamics able to confer motility to the flagellum \cite{winkler2020physics,schoeller2018flagellar}.

From a hydrodynamical point of view, the locomotion is possible because the viscous friction of the fluid along slender bodies is anisotropic \cite{fluidDyn_motility}; indeed, 
the hydrodynamic forces acting on slender bodies are predominantly anisotropic due to the different resistance encountered along and perpendicular to their length. This effect is embodied in the slender-body theory \cite{cox1970slender}, 
which simplifies the approach by focusing only on the body’s centerline, 
especially when viscous forces dominate (low Reynolds number). 
The implications of anisotropic hydrodynamic friction are essential to the comprehension in fields ranging from biological propulsion to the design of micro-scale devices. 
Although this issue is of great importance, it will not be addressed here.
\begin{figure}[h!]
\centering
\includegraphics[width=\columnwidth,clip=true]{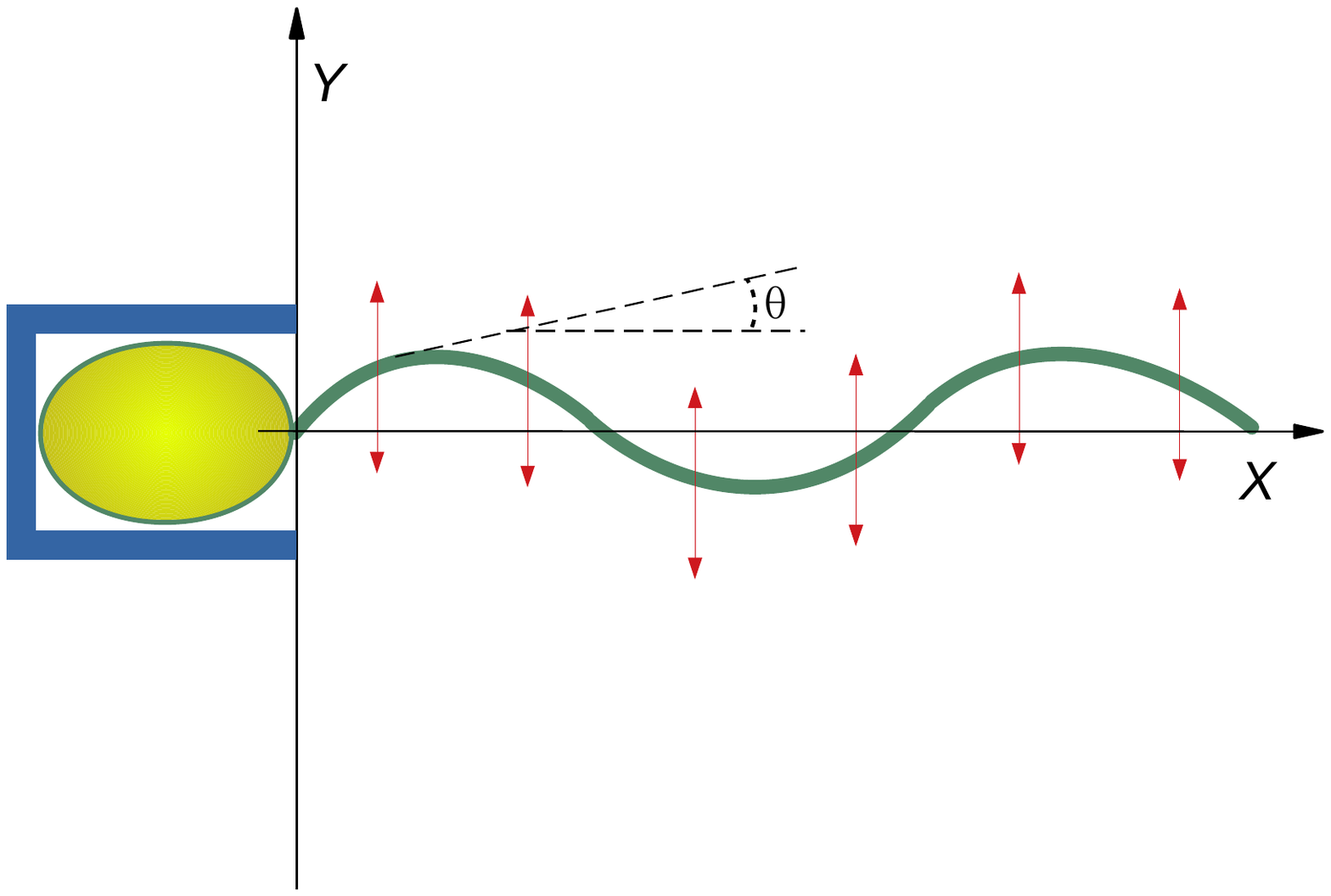}
\caption{\label{fig:bacter} Sketch of the two-dimensional geometry setup that we consider in this work. 
The head of the bacterium is blocked in a cage device (blue box), thus 
the flagellum is practically hinged to the origin and constrained to move on the plane XY. 
The red arrows represent the amplitude and direction of the 
traveling-wave forces driving the beating dynamics.}
\end{figure}
In this work, we study through simulations and theory a simplified, yet not trivial, version of an anchored flagellum depicted in Fig.\ref{fig:bacter} 
constrained to move in the 2D planar geometry and 
undergoing the effects of a travelling-wave active force which represents an idealized version of the coherent action of molecular motors 
during the beating dynamics.
In particular, we will analyze how spatially periodic conformational shapes of the flagellums can emerge from the interplay between its 
rigidity and the spatiotemporal periodicity of the activity \cite{artificialswim,spa-temp}. 
To maintain the analysis as simple as possible, we consider a monochromatic traveling-wave active force, characterized by a wavelength $\lambda$ 
(i.e. wavevector $k=2\pi/\lambda$) and an oscillation frequency $\omega$.
The study is performed at different $\omega$ and $k$ to show how these parameters affect the flagellum dynamical behavior.

The paper is structured as follows. In Section \ref{sec:model}, we introduce the computational two-dimensional model 
of the flagellum, along with its continuum theoretical version, 
which provides a useful framework for interpreting the simulation results. 
Section \ref{sec:results} discusses the dynamics of the flagellum, analyzing the computational results in terms of 
parameters $(\omega,k)$ that define the monochromatic active force. 
Finally, conclusions are drawn in Sec.\ref{sec:conclusion}.  

\section{Flagellum Models
\label{sec:model}}
\subsection{Computational Model}
In our approach, the flagellum is portrayed as a 2-dimensional chain of $N+1$ beads of mass $m$ connected by stiff elastic springs of finite extension $b$. As the first bead is anchored to the origin, the system cannot translate but can explore the full range of bending fluctuations.
The bending energy introduces a penalty whenever three consecutive beads lose their alignment. 
The potential energy of the system is defined by the function \cite{marques1997rigid,bark_BeadSpring,zeng2022dna} 
\begin{align}
	V(\mathbf{r}_0,\ldots,\mathbf{r}_N) &= 
	\frac{K}{2b^2} \sum_{n=1}^N \lx(|\mathbf{r}_n - \mathbf{r}_{n-1}| - b\rx)^2 \nonumber \\ 
	&+ \frac{B}{2b^2} \sum_{n=1}^{N-1}\lx|\mathbf{r}_{n+1} - 2 \mathbf{r}_n + \mathbf{r}_{n-1}\rx|^2.
	\label{eq:Vpot}
\end{align}
In this notation, spring $K$ and bending $B$ constants have the dimension of energy ($k_BT$); 
in the following, we will take $b=1$, $K\gg B$ ($K =100$), as we assume to work in a regime as similar as possible to an inextensible polymer. 
Even though a moderate degree of extensibility of the chain cannot be excluded \cite{lipowski_ExtSemiLex}, as small fluctuations of bond lengths $|\mathbf{r}_n - \mathbf{r}_{n-1}|$ are still possible around their expected value $b=1$.
 
We consider the overdamped regime of the system when coupled to a weak thermal bath, so the equation of motion of each bead is
 \begin{equation}
	 \gamma\,\dot{\mathbf{r}}_n = -\dfrac{\partial V}{\partial \mathbf{r}_n} + \mathbf{h}_n(t) + 
	 \sqrt{2\gamma T}\;\boldsymbol{\xi}_n(t)
	 \label{eq:motion}
 \end{equation}
 where $\boldsymbol{\xi}_n(t)$ is a Gaussian white noise of zero average and unitary variance, $T$ is the solvent (bath) temperature, while 
 $\mathbf{h}_n(t) = \lx(0,f_0 \sin(k\,n b - \omega t)\rx)$ is a deterministic space-modulated and time-periodic perturbation (active force) \cite{sec_harmo_beat} that, in our simplified scheme, confers to the flagellum a transversal beating dynamics.
 In the following, we set $f_0=B/20$ in all simulations and theoretical analysis to consider the active force as a relatively small perturbation.
 The active force $\mathbf{h}_n(t)$ competes with the bending rigidity in generating transversal undulatory fluctuations of the chain with a spatial modulation depending on the wave number $k$.  
 We choose $k=(2\pi/L)m$ where $L=N b$ is the length of the flagellum in terms of monomers and $m$ is a positive integer defining the number 
 of expected oscillations applied to the flagellum.
 This choice grants that the active force is spatially periodic over the polymer length $L$.

 In addition, we assume local isotropic dissipation with friction coefficient $\gamma$ and neglect long-range hydrodynamic interactions. 
 The role of $\gamma$ in the dynamics is to provide a time scale, i.e. the rate of energy dissipation due to the friction with the solvent, 
 and then it can be set $\gamma=1$ in the simulations without loss of generality. 
 Nevertheless, in the following theoretical analysis, we leave it indicated to keep track of the correct physical units of the observables;
 we also anticipate that the Gaussian noise will be neglected by the theory.
 This is because, in our simulations, the noise is completely overwhelmed by the forcing $\mathbf{h_n}$, i.e. $f_0^2 \gg \gamma T$.
 However, although small, its presence is necessary for running Brownian dynamics simulations \cite{huber2019brownian}. 

 We are interested in understanding how the interplay between bending rigidity and acting forcing selects the resulting 
 polymer conformations and in describing their relevant spatiotemporal features \cite{natali2020local,anchor_OscShear}.
 For this reason, we first carry out Brownian simulations \cite{huber2019brownian} of the chain through the Langevin Eq.\eqref{eq:motion}, then to better clarify the role of the perturbation parameters $(k,\omega)$, we support the computational results with a theoretical analysis based on the continuous model discussed below.
\begin{figure*}[t!]
\centering
\includegraphics[width=.92\textwidth,clip=true]{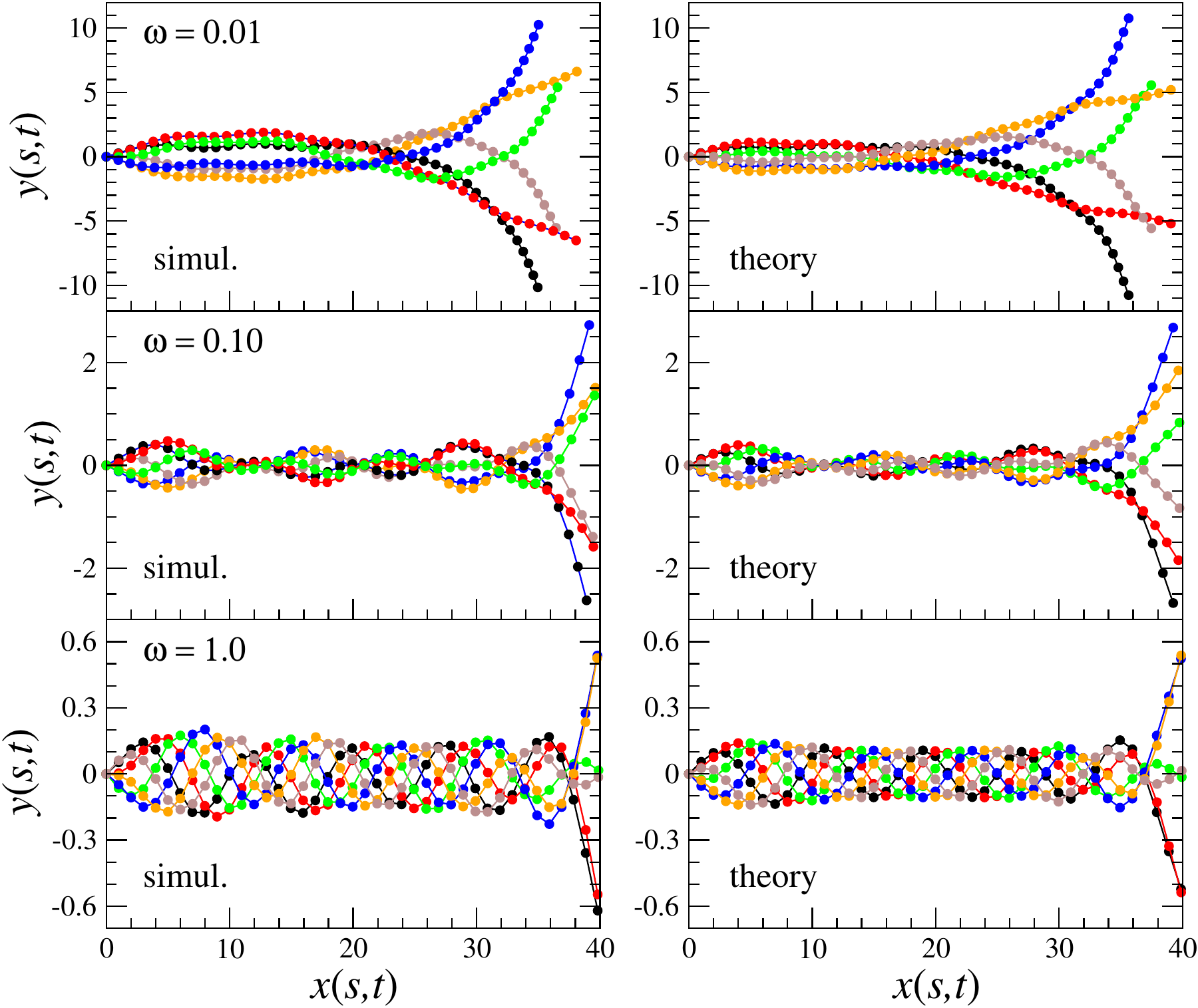}
\caption{Snapshots of the flagellum stationary dynamics (left panels) compared with the corresponding conformations obtained from the approximated 
 theory Eq.\eqref{eq:string}  (right panels), for a chain of length $L=Nb$ ($N=40$, $b=1$) and bending modulus $B=10$. 
 The different colors indicate six sampling times that are equally spaced within a force-cycle $2\pi/\omega$. 
 The comparison is done for active perturbations of wavevector $5(2\pi/L)$ and small, moderate, and high frequencies $\omega = 0.01,0.10,1.0)$.
\label{fig:flagcomp1}
}
\end{figure*}
 The left panels of Fig.\ref{fig:flagcomp1} and Fig.\ref{fig:flagcomp2} give the first qualitative idea about the beating 
 phenomenology of the system by showing several conformations assumed by the flagellum sampled in simulations over one period, $2\pi/\omega$, of the forcing.
 When $k=m (2\pi/L)$, the flagellum can sustain undulations of wavelength $\lambda=L/m$ as one can see from the number of bumps 
 along the conformations.
 The tail, instead, cannot follow the spatial periodicity due to the great moment of the external force on the free-end region making 
 the tail performs considerable excursions.

 Moreover, simulations indicate that at low frequencies, $\omega \lesssim 10^{-2}$, the flagellum oscillates over time but shows little 
 undulation.
 In contrast, at high frequencies, $\omega \gtrsim 1$, it becomes more spatially modulated and undergoes a "stiffening" transition, assuming a more rod-like conformation. 
 This occurs because its mechanical structure can no longer sustain and follow high-frequency modes. 
 The transition is evident from the "cigar-like" shape seen in the final left panels of Fig.\ref{fig:flagcomp1} and Fig.\ref{fig:flagcomp2}.
\begin{figure*}[t!]
\centering
\includegraphics[width=.92\textwidth]{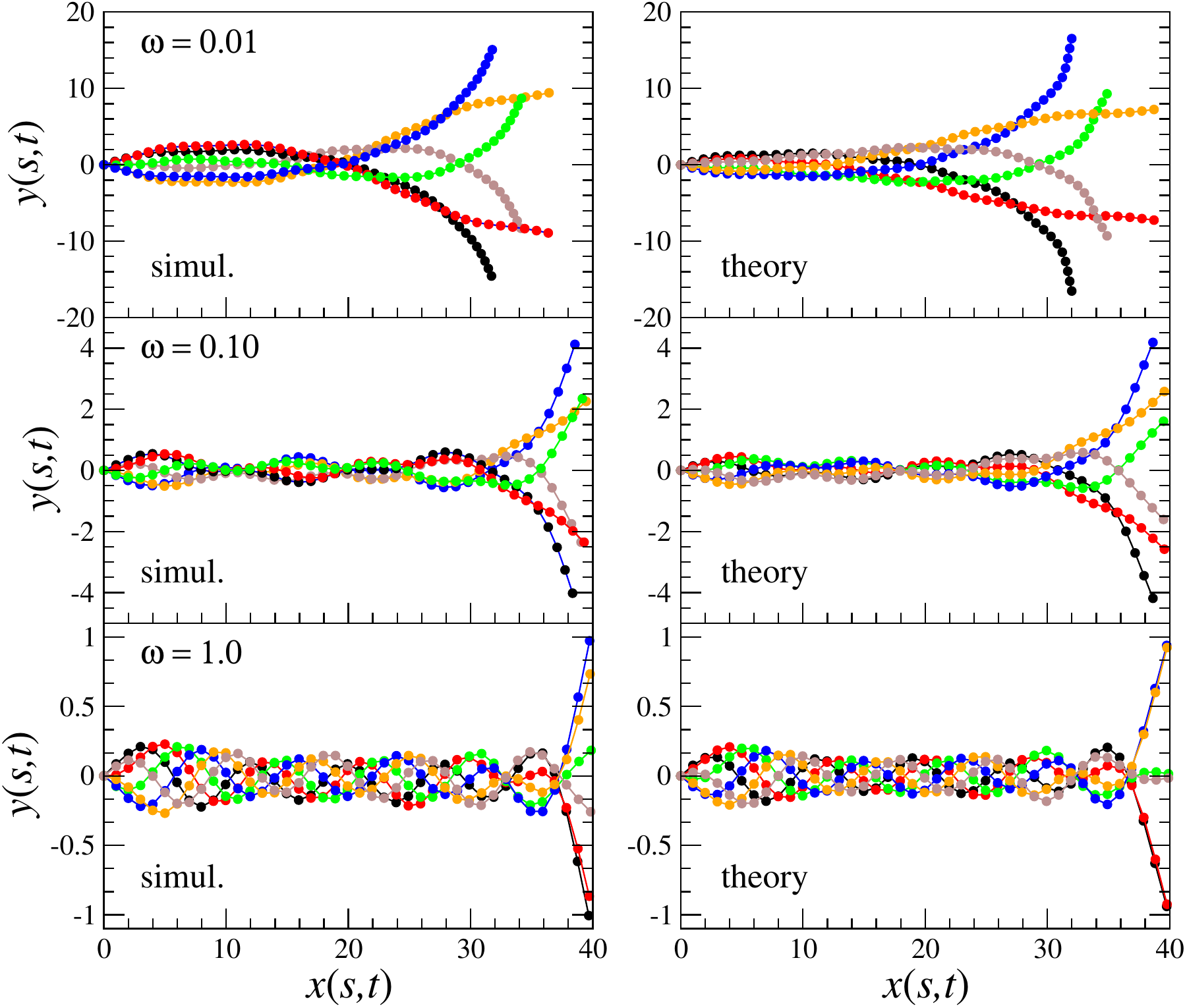}
\caption{Same conditions of Fig.\ref{fig:flagcomp1}, 
but higher bending stiffness, $B=20$.
\label{fig:flagcomp2}
}
\end{figure*}

 \subsection{Continuum Model 
 \label{sec:continuum}}
 The continuum formulation of model~\eqref{eq:motion} can be derived
 from the explicit expression of the internal forces obtained by direct differentiation of the potential \eqref{eq:Vpot}
 \begin{align}
	 -\dfrac{\partial V}{\partial \bfr_n} &= 
	     \frac{K}{b^2} 
	         [G\lx(\mathbf{r}_{n+1} -\mathbf{r}_n\rx) - G\lx(\mathbf{r}_n -\mathbf{r}_{n-1}\rx)] \nonumber \\
		     &-\frac{B}{b^2}\lx(\mathbf{r}_{n+2} -4\mathbf{r}_{n+1} +6\mathbf{r}_{n} -4\mathbf{r}_{n-1} +\mathbf{r}_{n-2} \rx)
		     \label{eq:rn}
 \end{align}
 with $\mathbf{G}(\mathbf{x})=(1-b/|\mathbf{x}|)\mathbf{x}$.
 Since the head of the chain is fixed while the tail is free,
 Eq.~\eqref{eq:rn} holds for all the bulk monomers except the boundaries (precisely beads of number $n=0,n=1,n=N-1,n=N$), we will discuss such conditions later.
 In the derivation of the continuum version of Eq.\eqref{eq:motion}, the flagellum is assimilated to a 2D-fluctuating string, hinged to the origin $\mathbf{r}(0)=0$, with elongation $2\nu = K/b$ and bending $\epsilon=B b$ stiffness. 
 The dimensional analysis suggests that $\epsilon/\nu$ has the dimension of a squared length, and more precisely, from Ref.\cite{winkler94}, $\epsilon/\nu = 2 \ell^2_p$, being $\ell_p$ the persistence length of the stiff polymer.
 As customary, we parameterize the curve with the arc length $bn \to s$ measured from the origin coinciding with the hinged point (see Fig.\ref{fig:bacter}).

 The equation of motion governing the evolution of the system's coordinate $\mathbf{r}(s,t) = x(s,t)\mathbf{e}_x + y(s,t)\mathbf{e}_y$ is straightforwardly derived from the monomers's Eq.\eqref{eq:rn} after performing the continuum limits $\mathbf{r}_n -\mathbf{r}_{n-1} \to b\; \partial \mathbf{r}(s)/\partial s$.
 In this way, the first term of Eq.\eqref{eq:rn} formally represents the discrete first derivative of the function $\mathbf{G}(\mathbf{x})$, whereas the last term is the discrete 4-th derivative of $\mathbf{r}(s)$ and neglecting the contribution of the Gaussian noise, this leads to
 \begin{align}
	 \gamma \dot{\mathbf{r}} &= 2\nu\frac{\partial}{\partial s}\left[g\lx(\lx|\partial_s \mathbf{r}\rx|\rx)\frac{\partial\mathbf{r}}{\partial s}\right] - \epsilon \frac{\partial^4\mathbf{r}}{\partial s^4} + \mathbf{h}(s,t)
	 \label{eq:string}       
 \end{align}
 where $g(x)=1-1/x$ and $\mathbf{h}(s,t) = \lx(0,f_0\sin(k s - \omega t)\rx)$.
 For consistency with the simulation setup, we have to assume: $\mathbf{h}(0,t) \equiv {\bf 0}$ because the external force on the hinged monomer is null.
 The boundary conditions for Eq.\eqref{eq:string} corresponding to a tethered-hinged polymer in the origin ($s = 0$), and free at the other end ($s = L$), 
 can be derived from Eqs.\eqref{eq:rn} by adding the ``virtual'' monomers $\mathbf{r}_{-1}, \mathbf{r}_{N+1}, \mathbf{r}_{N+2}$, 
 which must satisfy specific relations such that the equations of motion for the ``actual'' monomers $\mathbf{r}_1, \mathbf{r}_{N-1}, \mathbf{r}_{N}$
 can be treated as bulk Eqs.\eqref{eq:rn}. 
 This trick ensures the same description of the dynamics across the entire polymer chain, including the boundaries.
 Such conditions are, obviously, $\mathbf{r}_0=0$ and
 \begin{align*}
 &\mathbf{r}_{1} -2 \mathbf{r}_0 + \mathbf{r}_{-1} = 0\\
 &\mathbf{r}_{N+1} - 2\mathbf{r}_N + \mathbf{r}_{N-1} = 0\\
 &2 \nu g\lx(\frac{\lx|\mathbf{r}_{N+1}-\mathbf{r}_N\rx|}{b}\rx) \frac{\mathbf{r}_{N+1}-\mathbf{r}_N}{b} = \\
 &\epsilon \frac{\mathbf{r}_{N+2} - 3 \mathbf{r}_{N+1} + 3 \mathbf{r}_{N} - \mathbf{r}_{N-1}}{b^3}
 \end{align*}
 which, to the continuum, leads to $\mathbf{r}(0,t) = 0$ and
 \begin{align}
 &\dfrac{\partial^2\mathbf{r}}{\partial s^2}\bigg|_0 = 
 \dfrac{\partial^2\mathbf{r}}{\partial s^2}\bigg|_L = 0 \nonumber \\
 &2\nu g\lx(\lx|\partial_s \mathbf{r}\rx|\rx)\dfrac{\partial\mathbf{r}}{\partial s}\bigg|_L =
 \epsilon \dfrac{\partial^3\mathbf{r}}{\partial s^3}\bigg|_L
 \label{eq:bc}
 \end{align}
 The vanishing of the second derivative at $s=0$ and $s=L$ implies the absence of moments at the boundaries, while the last equation means no force at the free end $s=L$.
 In physical language, $g\lx(\lx|\partial_s\mathbf{r}\rx|\rx)$ corresponds to a line tension that guarantees the chain connectivity and its small
 stretchability \cite{ghosh_gov}, by introducing an energetic penalty paid to stretch (or contract) the filament beyond its rest length. 
 Mathematically speaking this term emerges by a variational principle on the WLC energy by imposing the constraint that the total chain length is left unaltered by the dynamical fluctuations \cite{harris1966polymer,frey_Rapid,ghosh_gov}.
 In our approach, the presence of a weak stretching contribution is simply justified by the simulation setup, for which the in-extensibility constraint is implemented only in a soft manner via stiff harmonic springs for the bonds \cite{bark_BeadSpring,panja2015efficient}. This, of course, does not prevent tiny fluctuations of the bond lengths around their fixed values $b=1$.
 Since our simulations are run under quasi-inextensible condition ($K \gg B$) we have $|r_n - r_{n-1}| \simeq b$ or, equivalently,  $\lx|\partial_s\mathbf{r}\rx| \simeq 1$ \cite{lipowski_ExtSemiLex}.
 It means that the factor $g\lx(\lx|\partial_s\mathbf{r}\rx|\rx)$ should be considered, if not vanishing, small.
 Moreover, the external force $\mathbf{h}$ that we implemented in our simulations tends to increase the distance between the beads; 
 this slight but distributed stretching of the chain ($\lx|\partial_s \mathbf{r}\rx|\gtrsim 1$) implies that the small value assumed by $g\lx(\lx|\partial_s\mathbf{r}\rx|\rx)$ is generally positive and weakly dependent on $s$.
 Nevertheless, trying to linearize directly Eq.\eqref{eq:string} by assuming a small and constant value of $\nu g\lx(\lx|\partial_s\mathbf{r}\rx|\rx) \to \nu_\mathrm{eff}$, leads to several inconsistencies, the worst of which is the collapse of the entire chain at the origin \cite{soda1973dynamics, aragon1985dynamics}.

 \subsection{Weakly Bending Approximation,
 \label{sec:WBA}}
 The weakly bending approximation (WBA) \cite{frey_Rapid,methods_Semiflex}
 is a convenient approach to recovering extensibility; 
 it assumes the flagellum undergoes only small deviations from a straight or rod-like conformation, 
 likewise, the active force is considered small enough to remain consistent with the 
 weakly bending regime.
 In our case, we see that WBA provides a satisfactory scheme for interpreting computational results. 
 As customary in WBA, a convenient parameterization of the flagellum centerline 
 is $\mathbf{r}(s,t) = \lx(s+u(s,t),y(s,t)\rx)$, where $u$ is a small longitudinal 
 deviation from $x = s$, with a small derivative as well, $\partial_s u(s)\ll 1$. 

 Such a parameterization is required to satisfy the inextensibility constraint
 \begin{equation}
	 |\partial_s\mathbf{r}|^2 = [1+\partial_s u(s,t)]^2 + [\partial_s y(s,t)]^2 = 1,
	 \label{eq:wbc}
 \end{equation}
 that, to the leading order, implies 
 \begin{equation}
	 \partial_s u(s,t) \simeq -\frac{1}{2}[\partial_s y(s,t)]^2,
	 \label{eq:dus}
 \end{equation}
 from which $u(s,t)$ is obtained by a direct integration, yielding
 \begin{equation}
	 x(s,t) = s - \frac{1}{2}\int_{0}^s d\xi\;
	 \bigg[\dfrac{\partial y(\xi,t)}{\partial \xi}\bigg]^2\,.
	 \label{eq:xsol}
 \end{equation} 
 In the WBA, the fluctuations $y(s,t)$ completely determines the shape of the flagellum, as 
 the variable $u(s,t)$, describing small longitudinal deformations, is enslaved to $y(s,t)$ through Eq.\eqref{eq:dus}. 
 To be consistent with the constraint, $g(\lx|\partial_s\mathbf{r}\rx|=1)=0$, the equation that the transversal 
 component $y(s,t)$ satisfies is
 \begin{equation}
	 \gamma \dot{y} = - 
	 \epsilon \frac{\partial^4 y}{\partial s^4} + f_0 \sin(k s-\omega t).
	 \label{eq:dotywoutnu}       
 \end{equation}
 Unfortunately, it provides solutions that deviate from numerical results, especially for the tail region, 
 as discussed in the section results, see Fig.\ref{fig:tail}.
 A decidedly better affinity is instead obtained by including a ``small'' stretching term leading to
 the effective equation
 \begin{align}
	 \label{eq:doty} 
	 &\gamma \dot{y} = 2\nu_\mathrm{eff} \frac{\partial^2y}{\partial s^2} - \epsilon \frac{\partial^4 y}{\partial s^4} + f_0 \sin(k s-\omega t) 
	 \\
	 &y(0,t)=0, \qquad  \lx.\frac{\partial^2 y }{\partial s^2}\rx|_0 =\lx.\frac{\partial^2 y }{\partial s^2}\rx|_L = 0 \nonumber 
	 \\
	 &\lx.2\nu_\mathrm{eff}\frac{\partial y }{\partial s}-\epsilon\frac{\partial^3 y }{\partial s^3}\rx|_L = 0 
	 \nonumber
 \end{align}
 with $\nu_\mathrm{eff}$ small and positive parameter that has to be determined.
 In the literature, $\nu_{\mathrm{eff}}$ is typically determined by imposing the average inextensibility condition of the flagellum \cite{harris1966polymer,actipoly}. 
 However, since we are already working within the WBA framework, which approximates this constraint, we need only to determine a reasonable value of $\nu$ from the simulation data (see next section).
 As a consequence, $\nu$ naturally turns out to be a function of the $\omega$ and $k$, $\nu_{\mathrm{eff}} = \nu_{\mathrm{eff}}(k,\omega)$ of the active force, like it always happens when a stiff polymer undergoes the action of an external driving (shear flow, pulling, or active random force) \cite{flex_in_Shear,wlc-tension,activesemiflex}. 

 In the rest of the paper, for simplicity, we use $\nu$ instead of $\nu_\mathrm{eff}$.

 Besides the need to reproduce numerical simulations, another guess for the presence of the stretching term stems from the observation 
 that by plugging the condition \eqref{eq:dus} into Eq.\eqref{eq:wbc} implies that 
 $\lx|\partial_s\mathbf{r}\rx|^2\simeq 1+(\partial_s y)^4/4$, thus $\lx|\partial_s\mathbf{r}\rx|$ is allowed to take on values slightly 
 larger than $1$.
 This is consistent with our expectation regarding the role of the term $g(\lx|\partial_s\mathbf{r}\rx|)$ in Eq.\eqref{eq:string}.

 Definitely, our analytical estimate of the flagellum conformations in the WBA is given by Eq.~\eqref{eq:doty} together with Eq.~\eqref{eq:xsol}.
 The solution to Eq.\eqref{eq:doty} can be obtained by an expansion 
 \cite{harris1966polymer,activesemiflex} 
 \begin{equation} \label{eq:sol}
	 y(s,t) = \sum_{n=1}^{\infty} y_n(t)\;\psi_n(s)
 \end{equation}
 in the orthogonal eigenmodes 
 \begin{equation}
	 \psi_n(s) = \dfrac{1}{\sqrt{W_n}} \bigg[
		 \dfrac{1}{\alpha_n^2}\dfrac{\sin(\alpha_n s)}{\sin(\alpha_n L)} + 
		 \dfrac{1}{\beta_n^2}\dfrac{\sinh(\beta_n s)}{\sinh(\beta_n L)} 
		 \bigg], 
		 \label{eq:funk}
 \end{equation}
 of the differential operator appearing in Eq.\eqref{eq:doty}, i.e.,
 \begin{align}
	 \epsilon \dfrac{d^4\psi_n(s)}{ds^4} - 2\nu \dfrac{d^2\psi_n(s)}{ds^2} = \lambda_n \psi_n(s), 
	 \label{eq:operator}
 \end{align}
 consistent with boundary conditions \eqref{eq:bc},  
 where $\lambda_n$ are the corresponding eigenvalues.
 Appendix \ref{app:eigen} reports the derivation of the eigenmodes and eigenvalues $\lambda_n$, 
 including the definition of $\alpha_n$ and $\beta_n$, 
 along with the formula of the normalization constants $W_n$.
 With a little abuse of language $\alpha_n$ or $\beta_n$ can be referred 
 to as ``quantum numbers'' of the eigenmodes.

 The amplitudes of each mode are independent and evolve according to the 
 following equation
 \begin{equation}
	 \gamma\;\dfrac{dy_n}{dt} = -\lambda_n y_n + b_n \cos(\omega t) - a_n \sin(\omega t)
	 \label{eq:dotQ}
 \end{equation}
 where
 \begin{equation}
	   a_n = f_0\langle\cos(ks),\psi_n(s)\rangle, \quad b_n = f_0\langle \sin(ks),\psi_n(s)\rangle,
 \end{equation}  
 are the coefficients representing the projection of the external force 
 $f_0\lx[\sin(ks)\cos(\omega t) - \cos(ks)\sin(\omega t)\rx]$ 
 on the eigenmodes; having defined the scalar product 
 $$
 \langle g(s), \psi_n(s) \rangle = \int_0^L ds\, g(s) \psi_n(s)\,.
 $$
 Their explicit expression is 
 \begin{align}
 a_n &= \dfrac{f_0}{\sqrt{W_n}}\bigg[
 \dfrac{\alpha_n \tan(\alpha_n L/2)}{\alpha_n^2 (\alpha_n^2 - k^2)} 
 + \dfrac{\beta_n\tanh(\beta_n L/2)}{\beta_n^2 (\beta_n^2 + k^2)}
 \bigg] 
\label{eq:a_n}
\\
\nonumber
\\
b_n &= \dfrac{f_0}{\sqrt{W_n}}\bigg[
 \dfrac{k}{\alpha_n^2 (\alpha_n^2 - k^2)} 
 - \dfrac{k}{\beta_n^2 (\beta_n^2 + k^2)}
 \bigg].
\label{eq:b_n}
\end{align}
In appendix \ref{app:mode}, we show that Eq.\eqref{eq:dotQ} is solved by the following (stationary) periodic solution:
\begin{equation}
 y_n(t) = 
 \dfrac{b_n \lambda_n + a_n (\gamma \omega)}{\lambda_n^2 + (\gamma \omega)^2}\cos(\omega t) +
 \dfrac{b_n (\gamma \omega) - a_n \lambda_n }{\lambda_n^2 + (\gamma\omega)^2}\sin(\omega t).
 \label{eq:yn}
\end{equation}
 We note that all coefficients appearing above, e.g. $a_n,b_n$ depend on the forcing parameters $(k,\omega)$, but -- to make the notation less burdensome, we have omitted such dependencies.

 After putting all the terms together and using the $\psi_n(s)$, we obtain the full solution of the y-profile of the flagellum that can be rearranged 
 into the very simple expression  
 \begin{equation}
 y(s,t) =  {\mathcal A}_{k,\omega}(s) \cos(\omega t) + {\mathcal B}_{k,\omega}(s) \sin(\omega t)
 \label{eq:ysol}
\end{equation}
 upon defining, for the sake of shorthand notation,  
\begin{align} 
{\mathcal  A}_{k,\omega}(s) &= \sum_{n=1}^{\infty}\psi_n(s)
\Big[
\dfrac{b_n \lambda_n + a_n(\gamma \omega)}{\lambda_n^2 + (\gamma \omega)^2}  
\Big] \nonumber \\
{\mathcal B}_{k,\omega}(s) &= \sum_{n=1}^{\infty} \psi_n(s)
\Big[
\dfrac{a_n (\gamma \omega) - a_n \lambda_n}{\lambda_n^2 + (\gamma \omega)^2}
\Big].
\label{eq:AandB}
\end{align}
 Here the indexes ``$k,\omega$'' recall the parametric dependence on the active force numbers.

 The amplitudes ${\mathcal A}_{k,\omega}(s)$ and ${\cal B}_{k,\omega}(s)$ play the role of a response of the system to a given forcing 
 with spatiotemporal periodicity $k,\omega$: high or low values of those amplitudes indicate if the traveling-wave perturbation is consistent with or is 
 attenuated by the bending rigidity of the flagellum. 
 Such amplitudes seem to be suppressed by increasing $\omega$, therefore in large $\omega$ regimes, the dynamics of the flagellum is expected to become noise-dominated and it loses the periodic behavior. In other terms, the flagellum response to high-frequency active forces resembles a low-pass filter 
 for which high-frequency perturbations are strongly attenuated, or in a more structural sense, this corresponds to a ``straightening'' crossover.

 It is also interesting to remark that the flagellum acts as a sort of ``spectral device'' which decomposes a ``monochromatic'' perturbation of 
 parameters $(k,\omega)$ into a superposition of responses 
 characterized by a set of numbers $\left\{\alpha_n,\lambda_n\right\}_{n=1}^{\infty}$ 
 where $\alpha_n$ (and $\beta_n$ that is strictly connected to it) plays the role of a generalized wave vector and $\lambda_n$ is the corresponding ``vibration'' frequency. 

\begin{figure}[h!]
\centering
\includegraphics[width=\columnwidth,clip=true]{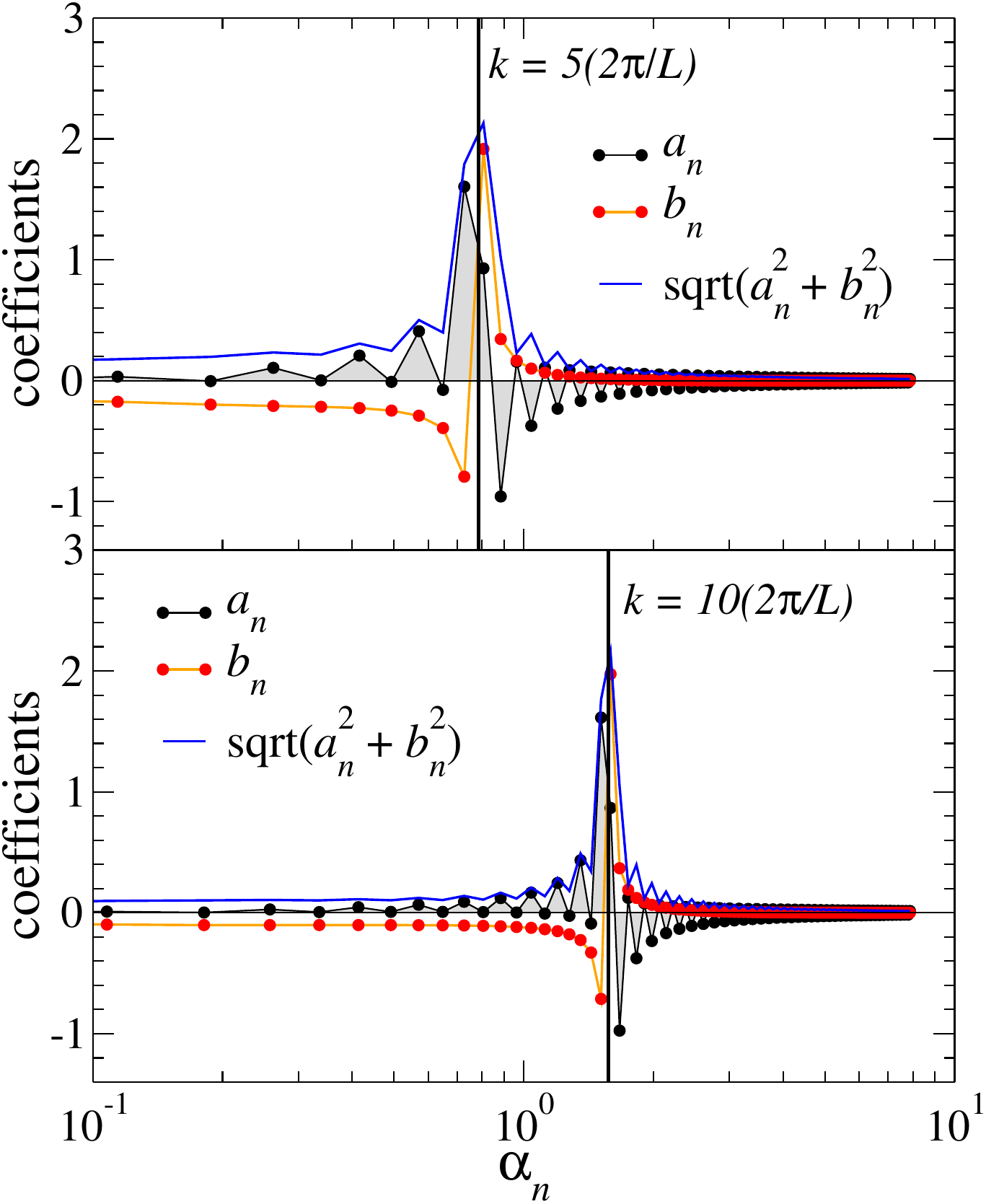}
\caption{\label{fig:alfa-kappa} 
Coefficients, Eqs.(\ref{eq:a_n},\ref{eq:b_n}) plotted versus $\alpha_n$, of the perturbation $f_0 \sin(k s-\omega t)$ expanded 
into the orthonormal eigenfunctions $\psi_n(s)$. 
The thick black vertical line marks the wave-number $k$ of the spatiotemporal forcing, $k=5(2\pi/L)$ and $k=10(2\pi/L)$.
 The blue line indicates the amplitude $\sqrt{a_n^2+b_n^2}$ of the $n$-th eigenmode. 
The picture resembles a ``spatial-resonance'' scenario because the maximal variation of coefficients occurs around $k$, 
even if the other contributions can not be considered negligible.
}
\end{figure}

 For visualizing the impact of the wave number, $k$, on the flagellum properties
 we plot, in Fig.\ref{fig:alfa-kappa}, the coefficients (\ref{eq:a_n},\ref{eq:b_n}) of the active-force expansion in eigenmodes as a function of $\alpha_n$.

 The coefficient $a_n$ (black dots) displays oscillations that intensify as $\alpha_n$ approaches $k$ (marked by the thick vertical line), 
 whereas $b_n$ (red dots) exhibits a growth towards $k$ without oscillations, undergoing a sign change when crossing $k$. 
 The behavior of $\sqrt{a_n^2 + b_n^2}$, which is a sort of composite envelope of $a_n$ and $b_n$, distinctly reveals a pronounced peak at $k$.

 These plots closely resemble a ``spatial resonance'' scenario, because the maximal variation of $a_n$ and $b_n$ occurs around a 
 neighborhood of $k$, even though the remaining contributions are not negligible.
 This suggests that the flagellum modes with $\alpha_n \simeq k$ are the most sensible to the 
 perturbation.

In the following section, we discuss the simulation results on the 
flagellum beating behavior by using the above continuum string theory as 
a reference and interpretative basis.  

\section{Results
\label{sec:results}}
We run simulations of a flagellum according to Eq.\eqref{eq:motion} for $N+1=41$ monomers (beads), the first of which is anchored to the origin, 
forming a hinged restrain, by numerically integrating Eq.\eqref{eq:motion} for each monomer forming the flagellum.
We select $K=100$, as a reasonable choice for granting inextensibility, and consider different bending modulus $B$ and active cycles, $2\pi/\omega$.
The time step is chosen to be of the order of $h\simeq 10^{-4}$ to avoid instabilities of the Brownian dynamics code (Euler scheme) 
within the whole explored range of parameters.
The flagellum is initialized by aligning all the monomers to the $x$-axis in their relaxed configuration: $x_n = n b$, ($n=0,\ldots,N$).
Main observables needed to characterize the flagellum dynamics were sampled and eventually averaged over a time $10 (2\pi/\omega)$, {\em i.e.} 
for ten forcing cycles, after discarding a transient of about $10^6$ time steps to allow the system's relaxation onto a robust stationary regime.

The left panels of Figs.\ref{fig:flagcomp1} and \ref{fig:flagcomp2} show snapshots of flagellum conformations 
from simulations for three values of $\omega$, with $B=10$ and $B=20$ respectively, while 
the right panels report the corresponding theoretical flagellum conformations obtained by solving Eq.\eqref{eq:doty} and 
using Eq.\eqref{eq:xsol} to reconstruct the $x$-coordinate in the WBA. 
The reasonable agreement between simulation and theory has been possible by adjusting the coefficient $\nu$ of 
the stretching term.
For calibrating $\nu$, we compared the motion of the flagellum tail (last monomer) with its theoretical prediction from Eq.\eqref{eq:doty}. 
\begin{figure}[h]
\centering
\includegraphics[width=0.97\columnwidth,clip=true]{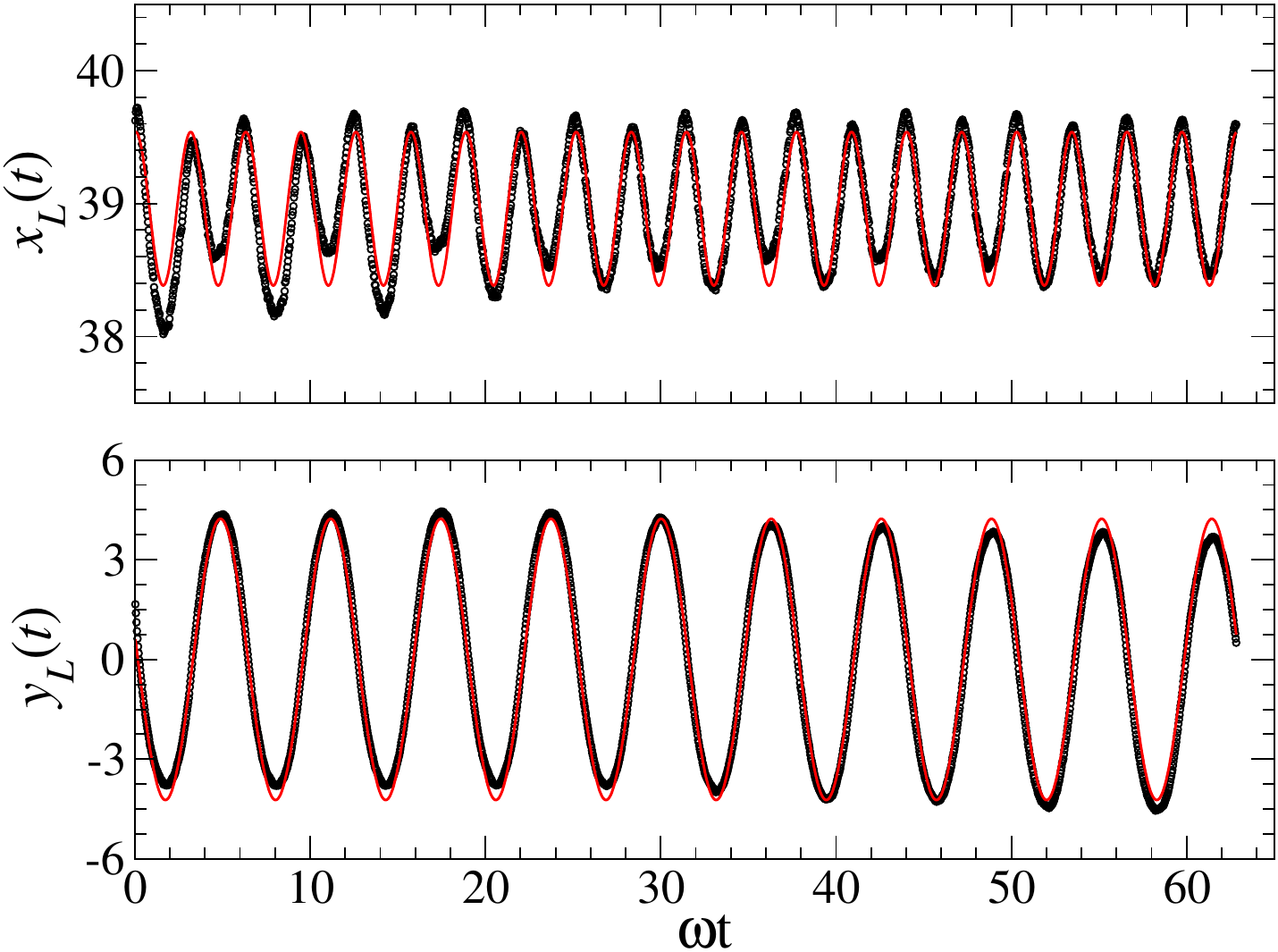}
\caption{Trajectory of the tail position $x_L(t), y_L(t)$ of 
a flagellum of $N+1=41$ beads (length $L=N b$), bending rigidity $B=20$ and under a active perturbation with 
$k=5, \omega=0.10$. 
Black dots are the simulated data and the red curves represent the fit with Eqs.(\ref{eq:fitx},\ref{eq:fity}). 
\label{fig:fitta}
}
\end{figure}
The beating oscillations of the tail observed in the simulations and displayed in Fig.\ref{fig:fitta} (black dots), can be described by the following simple evolutionary laws
\begin{align}
x(t) &= x_0 + a_x \cos(2\omega t) + b_x \sin(2\omega t) \label{eq:fitx}\\
y(t) &= a_y \cos(\omega t) + b_y \sin(\omega t).       \label{eq:fity}
\end{align}
As we will see in the following, the frequency doubling of the $x$ coordinate is a natural consequence of the quasi-inextensibility of the
flagellum. This phenomenon has also been observed in simulations of wall-anchored semiflexible polymers under oscillatory shear flow \cite{anchor_OscShear}.

\begin{figure*}[ht!]
\centering
\includegraphics[width=0.97\textwidth,clip=true]{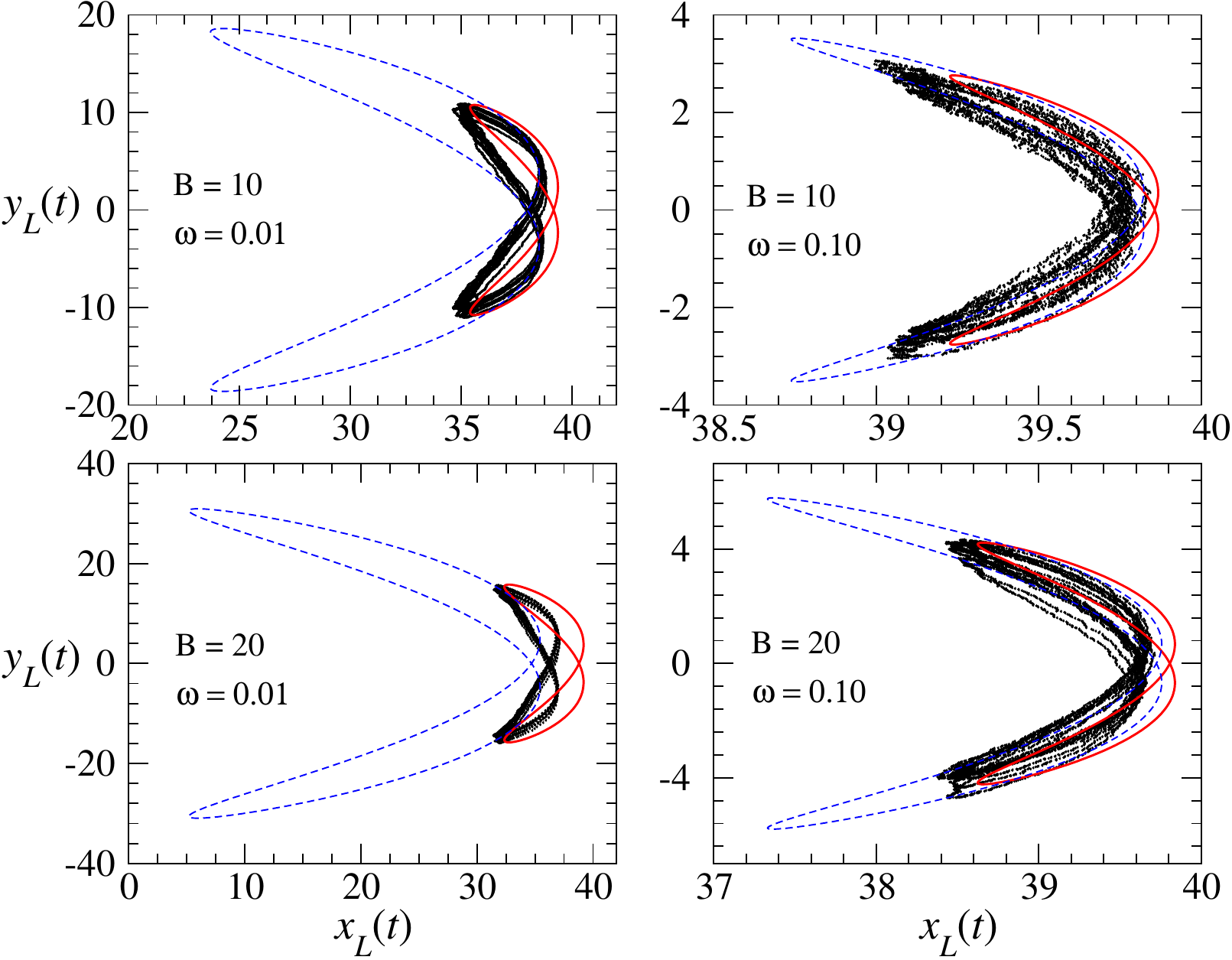}
\caption{
Trajectory described during the beating process by the position of the tail (last-monomer) of a flagellum 
with $N=40$ monomers of length $b=1$, for two
bending constants and two frequencies as indicated in the panels.
The motion recalls a bi-lobed Lissajous figure, the points are the simulation results, whereas the curves are the predicted trajectory described by Eqs.(\ref{eq:Lissa1},\ref{eq:Lissa2})  
The red-thick Lissajuous figures correspond to the optimal $\nu$ reported in Tab.1.  
For comparison, we also plot the dashed blue lines corresponding to the flagellum with $\nu=0$ to 
underscore the important role of the stretching contribution. 
\label{fig:tail}
}
\end{figure*}

The amplitude of the $y(t)$ signal is $F=\sqrt{a_y^2+b_y^2}$ whose numerical value can be obtained by a fitting procedure to
the simulation time series, see Fig.\ref{fig:fitta}.
The effective value of $\nu$ is such that $F^2 \simeq {\cA}^2(L) + {\cB}^2(L)$,
where
\begin{align}
{\cA}(L) &= \sum_{n=1}^{\infty}\frac{1}{\sqrt{W_n}}
    \bigg(\dfrac{1}{\alpha_n^2} + \dfrac{1}{\beta_n^2}\bigg)
   \dfrac{b_n \lambda_n + a_n (\gamma \omega)}{\lambda_n^2 + (\gamma \omega)^2}\,,\nonumber\\
{\cB}(L) &= \sum_{n=1}^{\infty}\frac{1}{\sqrt{W_n}}
\bigg(\dfrac{1}{\alpha_n^2} + \dfrac{1}{\beta_n^2}
\bigg)
   \dfrac{b_n(\gamma \omega) - a_n \lambda_n  }{\lambda_n^2 + (\gamma \omega)^2}\,.
   \label{eq:calAB}
\end{align}
are the amplitudes predicted by Eqs.\eqref{eq:AandB}, for shorthand notation we
dropped the indexes $\omega,k$.
The satisfactory numerical values of $\nu$ are reported in Tab.\ref{tab:nu_eff}.
\begin{table}[h!]
\centering
\caption{Table reporting the values of the estimated $\nu$
    obtained by comparing the amplitudes $F^2 = a_y^2 + b_y^2$ of the
    fitting Eq.\eqref{eq:fity} with the expected amplitudes
    ${\cal A}(L)^2 + {\cal B}(L)^2$ from the semiflexible flagellum theory.
} 
\label{tab:nu_eff}
\begin{tabular}{|c|c|c|c|c|}
\hline
  $\omega$ &  $0.01$ & $0.10$ & $1.0$ \\
    \hline
$B= 10$ & $\nu = 0.1696$ & $\nu = 0.1950$   & $\nu = 0.5632$ \\
    \hline
$B =20$ & $\nu = 0.3375$ & $\nu = 0.3710$  &  $\nu = 0.7150$\\
    \hline
\end{tabular}
\end{table}
Once we set an ``optimal'' value for $\nu$, we can compare the simulated trajectory of the flagellum tail (last monomer), describing a Lissajous-like figure in Fig.\ref{fig:tail} with the corresponding result obtained by using Eqs.(\ref{eq:xsol},\ref{eq:ysol}) at various values of $\omega$ and stiffness $B$
\begin{align}
\frac{x_L(t)}{L} &= 1 - \frac{1}{2L} \int_{0}^L ds [\partial_s y(s,t)]^2 \label{eq:L1} \\
\frac{y_L(t)}{L} &= \frac{\cA(L)}{L} \cos(\omega t) + \frac{\cB(L)}{L} \sin(\omega t)\,. \label{eq:L2}
\end{align}

The dashed lines in Fig.\ref{fig:tail} represent the theoretical Lissajous plots that by tuning $\nu$ converge and overlap with the 
simulation data.

The scattering (spread) observed in the simulation data is presumably 
due to the interplay of chaos and thermal noise which we do not investigate here. 
The reader can refer to Ref.\cite{chaos_semiflex} for a discussion of possible chaotic behaviors in anchored polymers 
driven by a localized oscillating force.
It is, however, clear that on passing from $\omega=0.01$ to $\omega =0.10$, the positions of the tail turn out to be more scattered, 
even if the thermal noise in the simulation is the same.
This is an intriguing role of the active forcing at high frequencies, which introduces a sort of ``stochastization effect''
that dominates over the small thermal noise, as shown in Eq.\eqref{eq:ysol}.

It is interesting to discuss how the bi-lobed Lissajous-like behavior of the tail emerges in terms of the continuum theory from Eqs.(\ref{eq:L1},\ref{eq:L2}). 
Figure \ref{fig:tail} shows that such Lissajous figures are characterized by a frequency ratio 2:1, corresponding to a {\em frequency doubling} of the $x$-motion: $\omega_x = 2\omega_y$. 
Indeed, the frequencies of a Lissajous' figure are known to satisfy the relationship 
$$
N_x\cdot \omega_x = N_y\cdot\omega_y\;,
$$
with $N_x$ and $N_y$ being the number of intersections of the curve with generic horizontal and vertical lines respectively and 
in our specific case $N_x=2$ and $N_y=4$, thus $\omega_x = 2 \omega_y$.

This ratio is expected as a straightforward consequence of the quasi-inextensibility of the chain, in fact 
by substituting Eq.\eqref{eq:L2} into Eq.\eqref{eq:L1}, and expressing formally the integrals as 
$L\langle\cdots\rangle$, we get 
\begin{align*}
\frac{x_L(t)}{L} &= S + \frac{Q-P}{4}\cos(2\omega t) - \frac{R}{2}\sin(2\omega t) \\
\frac{y_L(t)}{L} &= \frac{{\cal A}(L)}{L} \cos(\omega t) + \frac{{\cal B}(L)}{L} \sin(\omega t), 
\end{align*}
where we have defined 
$$
P = \langle [{\cal A}'(s)]^2\rangle,\quad Q=\langle [{\cal B}'(s)]^2\rangle, \quad  
R = \langle {\cal A}'(s){\cal B}'(s)\rangle, 
$$ 
with $S=1-(P+Q)/4$, and used the following trigonometric identities 
\begin{align*}
\cos^2(\omega t) = [1+\cos(2\omega t)]/2 \\
\sin^2(\omega t) = [1-\cos(2\omega t)]/2  
\end{align*} 
to make explicit the frequency doubling of the $x$-motion. 
The above equations, after simple manipulation, can be recast in the traditional form of 
Lissajous figures
\begin{subequations}
\begin{align}
x_L(t)/L &= S + S_x \cos(2\omega t + \Delta_x) \label{eq:Lissa1}\\
y_L(t)/L &= S_y \cos(\omega t + \Delta_y)\;,
\label{eq:Lissa2}
\end{align}
\end{subequations}
upon setting
$S_x\cos(\Delta_x) = (Q - P)/4$, 
$S_x\sin(\Delta_x) = R/2$
and $S_y\cos(\Delta_y) = {\cal A}(L)/L$, $S_y\sin(\Delta_y) = {\cal B}(L)/L$.
In conclusion, the tail's behavior can be mathematically explained by considering that the solution 
(\ref{eq:xsol},\ref{eq:ysol}) evaluated at $s=L$ results in a combination of $\sin(\omega t)$ and $\cos(\omega t)$ 
which can be rearranged in the form (\ref{eq:Lissa1},\ref{eq:Lissa2}). 
Moreover, the shape figure depends not only on the frequency (2:1) and amplitude ratio ($S_x$:$S_y$) but also 
on the phase shift, $\Delta = \Delta_y - \Delta_x$.

It's worth noting that each internal monomer of the flagellum also undergoes a similar type of Lissajous motion, 
albeit the greatest amplification is observed in the free tail.

\begin{figure}[ht]
\centering
\includegraphics[width=\columnwidth,clip=true]{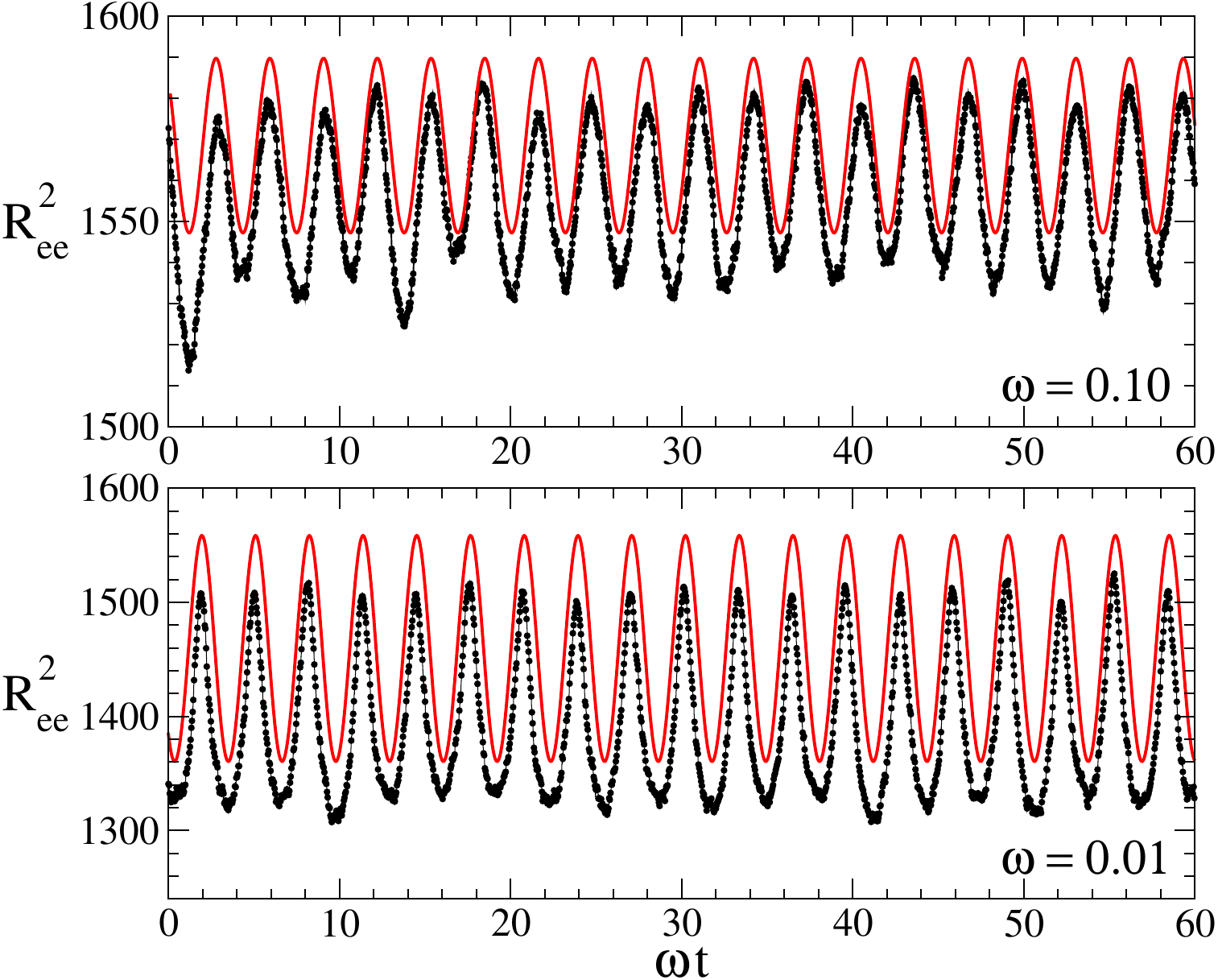}
\caption{Simulated time course (black) of the squared end-to-end distance $R^2_{ee}$ of a 
flagellum with $N+1=41$ beads each of diameter $b=1$, 
corresponding to a length $L=N b$. 
The chain has a bending rigidity strength $B=10$ and perturbation $k=5, \omega=0.01$.
and $\omega=0.10$. 
The simulations show a periodic behavior with a main frequency $\omega$ that is double the forcing frequency. 
The red curve represents the expected $R^2_{ee}$ obtained by the solution (\ref{eq:xsol},\ref{eq:xsol}).
\label{fig:Ree}
}
\end{figure}

Another quantity often used to characterize the dynamical response of the flagellum and its conformational 
properties is the end-to-end distance, which for the hinged system to the origin simply reads 
$$
R_\mathrm{ee}^2(t) = x^2(L,t) + y^2(L,t).
$$
Figure~\ref{fig:Ree} reports the time behavior of $R_{ee}(t)$ obtained from the simulations of a flagellum of $41$ beads (black curve) 
and compares it with the theoretical prediction (red curve) derived by squaring Eq.\eqref{eq:Lissa1} and Eq.\eqref{eq:Lissa2}.
The observable $R_\mathrm{ee}(t)$ exhibits a cyclical behavior with the active-force period $2\pi/\omega$ which is well 
reproduced by the theoretical curve (red) once the free parameter $\nu$ in Eq.\eqref{eq:string} is properly set, as reported in 
Tab.\ref{tab:nu_eff}.      

\subsection{Spatial Modulation}
So far, we have analyzed the influence of the force cycle, defined by $\omega$, on the flagellum beating dynamics, 
now, we would like to focus on the role of $k$ in shaping the spatial conformation of the flagellum. 
To check if the spatial modulation induced by the perturbation is sustained 
by the filament dynamics, we computed the bond-bond correlation 
starting from the hinged point, $x_0=0, y_0=0$,
\begin{equation}
C_b(n) =  \overline{\mathbf{u}_1 \cdot \mathbf{u}_n} \qquad n=1,\ldots,N,
\label{eq:corr_n}
\end{equation}
with $\mathbf{u}_n = \mathbf{r}_{n+1} - \mathbf{r}_{n}$ being the bond vector connecting the positions of two consecutive beads. 
The notation $\overline{f(t)}$ indicates the time average, which in our case, coincides with the average over a period, $2\pi/\omega$, of the active force.
\begin{figure}[h!]
\centering
\includegraphics[width=.92\columnwidth,clip=true]{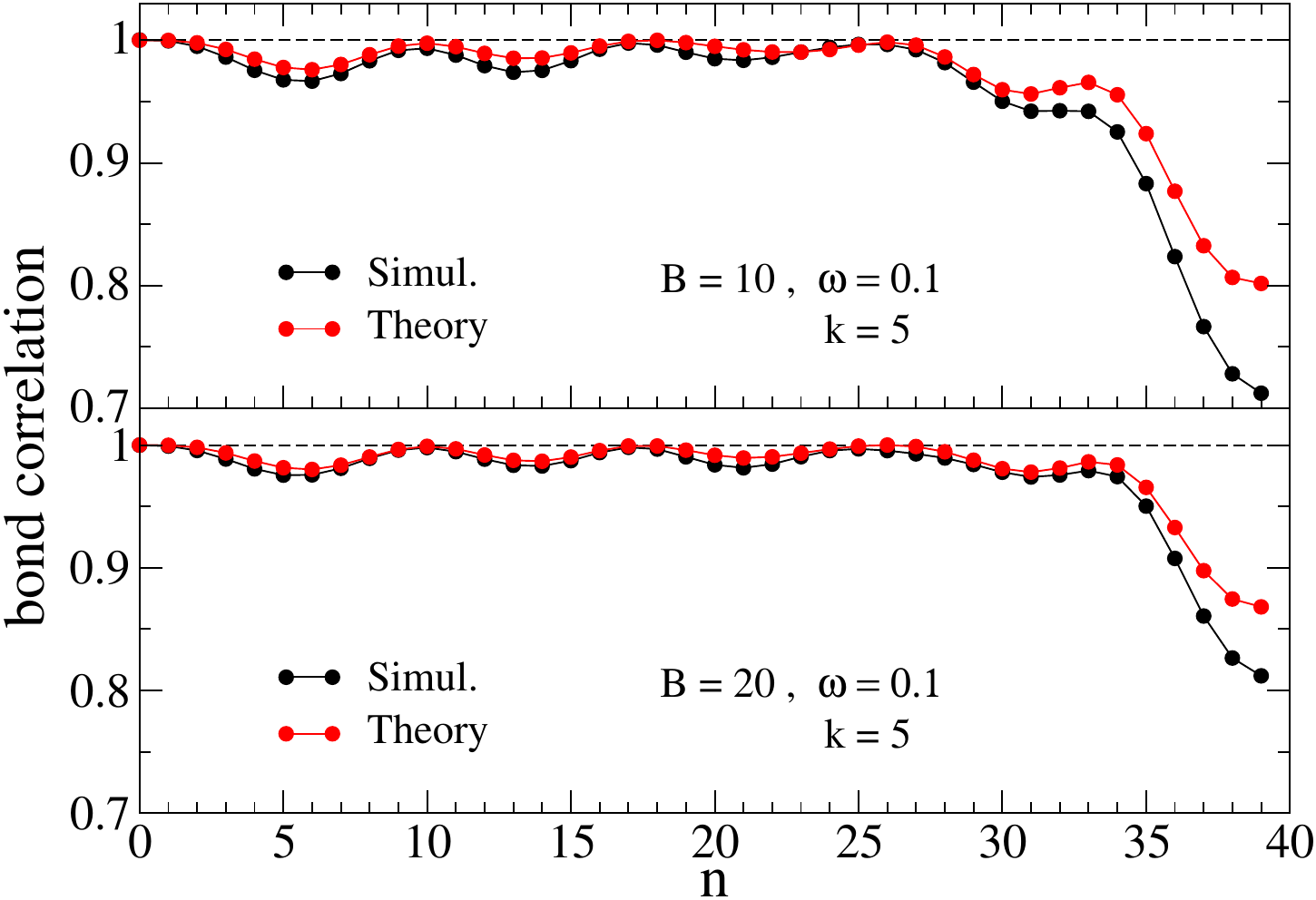}
\caption{
Bond-bond correlation $C(n)$ for filaments of $N=40$ monomers, with bending stiffness $B=10$ and $B=20$, both having the same $\omega = 0.10$. 
Black dots represent simulation data, while red dots represent theoretical predictions. 
Both the simulation and theoretical plots reveal spatial modulation with five peaks, consistent with the active-force wave number $k = 5(2\pi/L)$, where $L = N b$.
The average is taken over a long equilibrium simulation run at temperature $T = 0.001$.
The red plots, drawn from Eq.\eqref{eq:corrs}, show a pattern of peaks similar to the simulation data.
\label{fig:cbond}
}
\end{figure}
Figure \ref{fig:cbond} presents the simulated data for $C_b(n)$ (black points) averaged over a long equilibrium run at low temperature $T=0.001$, 
corresponding to a flagellum of bending stiffness $B=10$, (upper panel) and $B=20$ (lower panel) both perturbed with frequency $\omega=0.10$.
$C_b(n)$ exhibits five peaks, consistent with the oscillations imposed by the active-force wave number, $k=5(2\pi/L)$, indicating 
that the chain's stiffness can support the external modulation.

Simulations also show that the bond orientation becomes more correlated as the bending stiffness increases, this is expected since the persistence length 
is higher for $B=20$.
The faster decay of the correlation for $B=10$ is also indicative of weaker persistence, and a ``visual estimate'' of the persistence length of the 
flagellum can be obtained from the onset of a large deviation of the tail region from the horizontal dashed line.

A theoretical prediction of bond-bond correlation behavior can be obtained from the continuum model by using the following formula 
\begin{equation}
    C_b(s) = b^2 \overline{\frac{\partial \mathbf{r}(s,t)}{\partial s}  \cdot \frac{\partial \mathbf{r}(0,t)}{\partial s} }
   \label{eq:corr_s} 
\end{equation} 
in which $b\,\partial_s \mathbf{r}(s,t)$ is the continuum analogue of a bond vector in Eq.\eqref{eq:corr_n}.
The overbar denotes the time average over a force cycle ($2\pi/\omega$) and we recall that, throughout the text, we set $b=1$ to eliminate inessential parameters.
Figure \ref{fig:cbond} presents the simulated data for $C_b(n)$ (black points) averaged over a long equilibrium run at low temperature $T=0.001$, 
corresponding to a flagellum of bending stiffness $B=10$, (upper panel) and $B=20$ (lower panel) both perturbed with frequency $\omega=0.10$.
$C_b(n)$ exhibits five peaks, consistent with the oscillations imposed by the active-force wave number, $k=5(2\pi/L)$, indicating 
that the chain's stiffness can support the external modulation.

In the WBA, the bond vector can be written as 
\begin{align*}
\frac{\partial\mathbf{r}}{\partial s} &= 
\lx(1 - \frac{1}{2}
\lx(\frac{\partial y}{\partial s}\rx)^2,\frac{\partial y}{\partial s} \rx)
\end{align*}
therefore, $C_b(s)$ to the leading order becomes
\begin{align*}
C_b(s) &\simeq 1 - 
\frac{1}{2}\overline{\lx[\frac{\partial y(s,t)}{\partial s}\rx]^2} - \frac{1}{2}\overline{\lx[\frac{\partial y(0,t)}{\partial s}\rx]^2} + \\
&-\overline{\frac{\partial y(s,t)}{\partial s}\frac{\partial y(0,t)}{\partial s}}.
\end{align*}

By using Eq.\eqref{eq:ysol} and the average properties of trigonometric functions over a period, after simple algebraic manipulations, $C(s)$ can be recast to the form  
\begin{equation}
C_b(s) = 1 - \frac{1}{4}
\bigg([{\cal A}'(s) - {\cal A}'(0)]^2 + [{\cal B}'(s) - {\cal B}'(0)]^2
\bigg)
\label{eq:corrs}
\end{equation}
where ${\cal A}$ and ${\cal B}$ are the expressions in Eq.\eqref{eq:calAB}.
Notice that correctly $C_b(s=0) = b^2 = 1$, as expected from the ``quasi-inexensibility'' of the flagellum bonds.

For a comparison, Fig.\ref{fig:cbond} also displays the corresponding quantity $C_b(s)$ derived from the continuum model, Eq.\ref{eq:corrs} (red points).  
Although a precise quantitative agreement between $C_b(n)$ and $C_b(s)$ is lacking especially in the tail region where the application of WBA is 
questionable, it is noteworthy that they exhibit five similar oscillations driven by the active forcing modulation, $k=5(2\pi/L)$.

This suggests that the qualitative structure of the beating dynamics remains consistent when passing from the discrete to the continuum models.

\section{Conclusions
\label{sec:conclusion}}
The main purpose of this study was to investigate the beating dynamics of 
flagella through a combination of semiflexible slender structure and the 
active forces driving its temporal evolution across different $\omega,k$ 
regimes.

In particular, we developed an effective theory, that upon ``tuning'' the longitudinal stretching term, can explain (at least qualitatively) some 
important features of the beating dynamics observed in the simulation, such as the bond-bond spatial correlation and temporal oscillation of the 
flagellum tail which, in a certain frequency range of the active force, describes Lissajous' figures of 2:1 frequency ratio.
The theory clearly explains that such a 2:1 frequency ratio arises from a {\em frequency doubling} of the $x$-oscillation, which in turn, is 
a natural consequence of the chain inextensibility.

Furthermore, the theoretical analysis sheds light on the ranges of $\omega$ and $k$ of the active force required to confer a prescribed spatial 
modulation to the flagellum and producing a significant effect on its beating dynamics.
Indeed, as explained in Sec.~\ref{sec:continuum}, for high-frequencies forcing ($\omega\gg1$) the amplitude of periodic 
modulations of $y(s,t)$ become largely suppressed, and thermal fluctuations dominate the flagellum dynamics.  

Additionally, the wavelength, $2\pi/k$, of the active forcing must be consistent with the persistence length, $l_p$, of the flagellum,
as high levels of spatial modulation cannot be energetically sustained by a system with a given bending stiffness, see Appendix \ref{app:simple} for a less qualitative argument.
Such a result could play a key role in designing bio-inspired self-propelled engines. 

Although the modeling of active forcing is not fully realistic from a biological point of view, we expect that the theory 
could be generalized to describe different scenarios, including cases where the active force has an ``internal'' origin by modeling 
the action of molecular motors distributed along the flagella.
Thus, this work constitutes a first step towards more realistic modeling of active bio-filaments giving insight into the mechanism 
responsible for their dynamics and effective propulsion.

\section*{Data availability}
The source files of Figures 2, 3, 4, 5, 6, 7, and 8 of the manuscript are 
available at https://github.com/cecconif/flagellum-wlc-data.
Each file can be opened using the graphical software Grace, which can be found
and downloaded from the site https://plasma-gate.weizmann.ac.il/Grace/.
Such files also contain, at the end, the raw data used to generate the figures.

\section*{Acknowledgements}
The authors are grateful to A.Cavagna for stimulating discussions 
and his precious remarks and suggestions.


\bibliography{litera}

\appendix
\section{Derivation of the Eigenmodes \label{app:eigen}}
In this appendix, we derive the eigenmodes which are solutions of the
equation
\begin{equation}
\epsilon \dfrac{d^4\psi_n(s)}{ds^4} - 2\nu \dfrac{d^2\psi_n(s)}{ds^2} = \lambda_n \psi_n(s)
\label{eq:autofunk}
\end{equation}
satisfying the boundary conditions:
\begin{align*}
& \psi_n(0) = 0, \qquad \dfrac{\partial^2\psi_n(s)}{\partial s^2}\bigg|_0 = 0 \\
&\dfrac{\partial^2\psi_n(s)}{\partial s^2}\bigg|_L=0, \qquad
2\nu \dfrac{\partial\psi_n(s)}{\partial s}\bigg|_L =
\epsilon \dfrac{\partial^3\psi_n(s)}{\partial s^3}\bigg|_L
\end{align*}
Since Eq.\eqref{eq:autofunk} is linear, its solution requires solving the associated characteristic polynomial
$$
\epsilon\,\mu^4 -2\nu\,\mu^2 - \lambda_n = 0,
$$
the roots of which are, complex $\mu=\pm i\alpha$ and real $\mu=\pm\beta$, where
$$
\alpha^2 = \sqrt{\bigg(\frac{\nu}{\epsilon}\bigg)^2 + \frac{\lambda}{\epsilon}} -
\frac{\nu}{\epsilon}\,,
\quad
\beta^2 = \sqrt{\bigg(\frac{\nu}{\epsilon}\bigg)^2 + \frac{\lambda}{\epsilon}} +
\frac{\nu}{\epsilon}\,,
$$
then, it is straightforward to obtain the following algebraic identities
\begin{align}
&\beta^2-\alpha^2 = 2\nu/\epsilon \\
&\lambda = \epsilon \alpha^4 +2\nu \alpha^2 = \epsilon \beta^4 -2\nu \beta^2.
\end{align}
Then the general solution of Eq.\eqref{eq:autofunk} can be written as
\begin{equation}
\psi(s) = A \sin(\alpha s) + B\sinh(\beta s)+
          C \cos(\alpha s) + D\cosh(\beta s);
\label{eq:psigen}
\end{equation}
the coefficients $A, B, C, D$ and the eigenvalues $\lambda$ are determined by imposing the four boundary conditions plus the normalization.
The condition at $s=0$ obviously implies that $C=D=0$, while the condition in $s=L$
implies
\begin{align*}
\alpha^2\sin(\alpha L) A & - \beta\sinh(\beta L)  B = 0 & \\
\beta   \cos(\alpha L) A & - \alpha\cosh(\beta L) B = 0 &,
\end{align*}
to simplify, we used the identity $\epsilon(\beta^2-\alpha^2) = 2\nu$.
In matrix form, the boundary conditions lead to the following linear systems
$$
\begin{bmatrix}
          0 &  0 &     1     &   1       \\
          0 &  0 & -\alpha^2 & \beta^2   \\
\alpha^2\sin(\alpha L) & - \beta^2\sinh(\beta L) & 0 & 0  \\
\beta\cos(\alpha L)    & - \alpha\cosh(\beta L) & 0 & 0
 \end{bmatrix}
\begin{bmatrix}
A\\B\\C\\D
\end{bmatrix}  =
\begin{bmatrix}
0\\0\\0\\0
\end{bmatrix}.
$$
To exclude the nontrivial solution $A=B=D=0$,
the determinant of the coefficient matrix should be zero,
therefore we have the condition
$
\alpha^3\sin(\alpha L)\cosh(\beta L) = \beta^3 \cos(\alpha L)\sinh(\beta L)
$ that can be recast to
\begin{equation}
\alpha^3\tan(\alpha L) = \beta^3 \tanh(\beta L)
\label{eq:lambda}
\end{equation}

After simple algebraic manipulations, we obtain from Eq.\eqref{eq:psigen}
the final expression of the eigenfunctions,
\begin{equation}
\psi_n(s) = W_n \bigg[
\dfrac{1}{\alpha_n^2}\dfrac{\sin(\alpha_n s)}{\sin(\alpha_n L)} +
\dfrac{1}{\beta_n^2}\dfrac{\sinh(\beta_n s)}{\sinh(\beta_n L)}
\bigg],
\label{eq:psi_n}
\end{equation}
$W_n$ being a normalization constant such that
$$
\int_{0}^L ds\;\psi_n^2(s) = 1,
$$
then using the trigonometric identities, $\sin^{-2}(p) = \cot^2(p)+1$, $\sinh^{-2}(p) = \coth^2(p)-1$, and Eq.\eqref{eq:lambda}, this can be  expressed as
\begin{align}
 W_n^{-2} &= L \frac{2\nu^2+\epsilon\lambda_n}{\lambda^2_n} \nonumber \\
            &+ \frac{\coth(\beta_n L)}{\beta_n^3}
            \bigg(\frac{3 \nu}{\lambda_n}
            -\frac{L\nu \coth(\beta_n L)}{\epsilon \beta_n^3} \bigg)
\end{align}

\begin{figure}[h!]
\centering
\includegraphics[width=.9\columnwidth,clip=true]{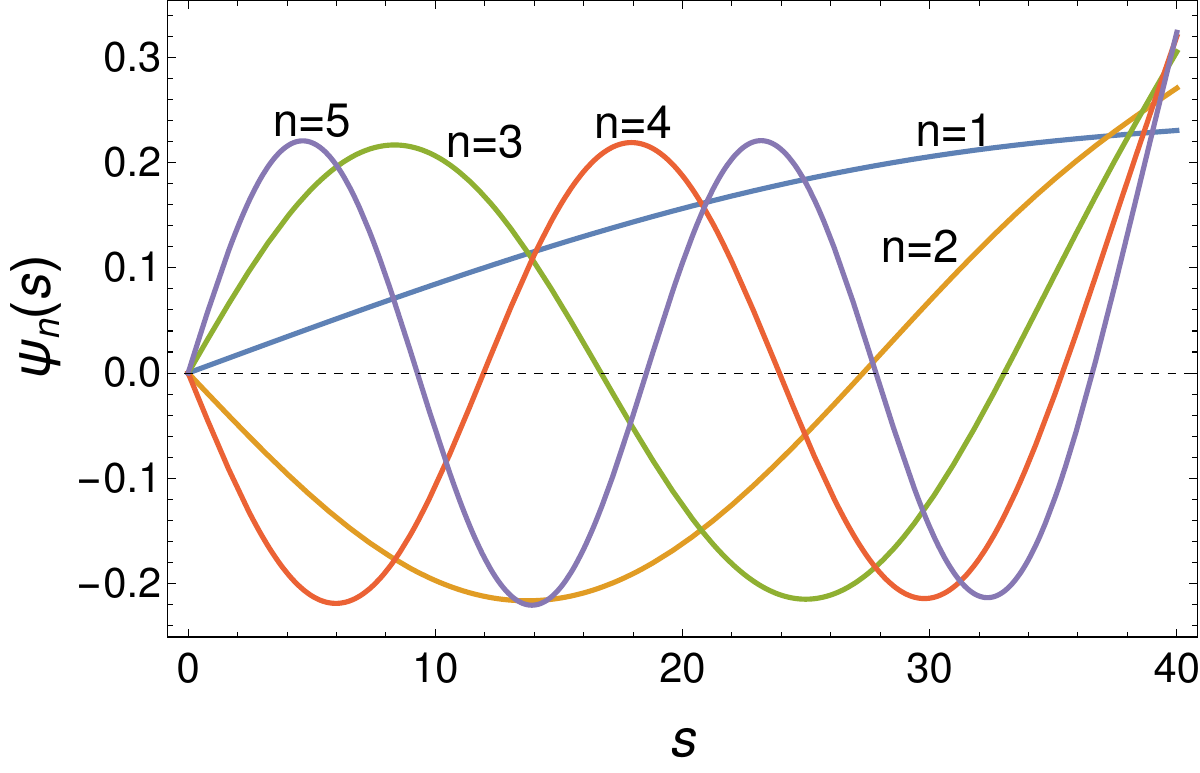}
\caption{First five eigenfunctions $\psi_n(s)$ as a function of the 
arclength coordinate, $s$. 
Oscillations are associated with the generalized wavenumber $\alpha_n$, which must be compared with the wavenumber $k$ of the active force modulation.  
\label{fig:eigf}
}
\end{figure}

\section{Solution of the mode amplitude equation
\label{app:mode}}
We derive the evolution of the n-th mode amplitude, $y_n(t)$, that obeys the equation 
\begin{equation}
\gamma\;\dfrac{dy_n}{dt}  + \lambda_n y_n = 
a_n \cos(\omega t) + b_n \sin(\omega t)
\label{eq:moveY}
\end{equation}
The solution of this first-order equation in the stationary regime is
$$
y_n(t) = \dfrac{e^{-\lambda t/\gamma}}{\gamma} \int_{-\infty}^{t}\!\!dz\;e^{\lambda z/\gamma}[a_n \cos(\omega z) + b_n \sin(\omega z)\big] 
$$
after a change of variable $t-z \to z$
$$
y_n(t) = \int_{0}^{\infty}
\!\!dz\;\dfrac{e^{-\lambda z/\gamma}}{\gamma} \Big\{a_n \cos[\omega(t- z)] + b_n \sin[\omega(t- z)]\Big\} 
$$
and expanding $\cos[\omega(t- z)]$ and $\sin[\omega(t- z)]$, we obtain
\begin{equation*}
y_n(t) = (a_n Z_1 - b_n Z_2)\cos(\omega t) + (a_n Z_2+ b_n Z_2)\sin(\omega t)
\end{equation*}
where
$$
Z_1 = \frac{1}{\gamma} \int_{0}^{\infty}
\!\!dt\;e^{-\lambda t/\gamma}\cos(\omega t) = 
\dfrac{\lambda}{\lambda^2 + (\gamma \omega)^2}
$$
$$
Z_2 = \frac{1}{\gamma} \int_{0}^{\infty}
\!\!dt\;e^{-\lambda t/\gamma}\sin(\omega t) = 
\dfrac{\gamma \omega}{\lambda^2 + (\gamma\omega)^2}.
$$
Thus re-arranging, we can write
$$
y_n(t) = \dfrac{a_n \lambda - b_n (\gamma \omega)}{\lambda^2 + (\gamma \omega)^2}\cos(\omega t) +        
\dfrac{a_n (\gamma \omega)+ b_n\lambda}{\lambda^2 + (\gamma \omega)^2}\sin(\omega t)\,,
$$
\\
which is the searched solution, Eq.\eqref{eq:yn} of the main text.

The same expression could have been obtained, by substituting the test function (similar to the known term)
$y_n(t) = C_1 \cos(\omega t) + C_2 \sin(\omega t)$ into Eq.\eqref{eq:moveY} and choosing the coefficients $C_1,C_2$ to make both members equal.

\section{A simple criterium \label{app:simple}}
In this appendix, we discuss a simple energetic argument that justifies
the physical limits of the flagellum response.

The argument compares the energy of a perturbation mode
and the energy of the flagellum solution \eqref{eq:sol}, using the continuum energy formula \cite{frey_Rapid}
\begin{equation}
E = \int_{0}^L ds \Big[
\dfrac{\epsilon}{2}
\Big(\dfrac{\partial^2\mathbf{r}}{\partial s^2}\Big)^2 +
\nu \Big(\dfrac{\partial\mathbf{r}}{\partial s}\Big)^2
\Big]
\label{eq:Eng}
\end{equation}
The active perturbation reads
$$
F(s,t) = 2\dfrac{f_0}{L} \sin(k s - \omega t)
$$
where, the pre-factor $2 f_0/L$ is the necessary scale to make the amplitude of $F(s,t)$ and the flagellum longitudinal
oscillation $y(s,t)$ of the same order.
When substituted into Eq.\eqref{eq:Eng}, $F(s,t)$ requires an energetic cost
$$
E(k) = \dfrac{f_0^2}{4}(\epsilon k^4 + 2\nu k^2)\;.
$$
This cost has to be compared with the energy of the flagellum solution, which can be computed using its expansion in eigenmodes
\eqref{eq:sol}
$$
E_f = \dfrac{f_0^2}{4}
\sum_{n=1}^{\infty} \dfrac{\overline{a}^2_n + \overline{b}_n^2}{\lambda_n^2 + (\gamma \omega)^2}
\lambda_n\;,
$$
note that, with respect to the expressions \eqref{eq:a_n} and \eqref{eq:b_n}, the forcing amplitude, $f_0$, has been factored out,
thus we obtain
\begin{equation}
\epsilon k^4 + 2\nu k^2 \simeq
\sum_{n=1}^{\infty} \dfrac{\overline{a}^2_n + \overline{b}_n^2}{\lambda_n^2 + (\gamma \omega)^2}.
\end{equation}
This equation suggests a criterion that determines the range of $k$ and $\omega$.
Specifically, once $k$ is chosen, the equation determines a range of feasible $\omega$ values around the solution of the equation.
Conversely, if $\omega$ is assigned, a range of feasible $k$ values can be derived from the equation.

In other terms, if $E(k)$ is too large due to a high value of $k$, we introduce too much energy into the system, which must be balanced by choosing a small $\omega$, and vice-versa, if $k$ is too small, it requires a very large $\omega$.

This situation is also critical because, at high frequencies, the system dynamics are significantly attenuated, leading to a decoupling between the flagellum's dynamics and the fast oscillating perturbation: in this regime, the flagellum perceives this rapid oscillation as additional ``noise''.

\end{document}